\documentclass{PoS}

\usepackage{multirow}
\def \ketv #1>{\mbox{$|{#1}\rangle$}} 
\def \brav #1|{\mbox{$\langle {#1}|$}}
\def \mate<#1|#2|#3>{\mbox{$\langle {#1}|\,{#2}\,|{#3}\rangle$}}

\title{Flavor structure of the baryon-baryon interaction from lattice QCD}

\ShortTitle{Flavor structure of the baryon-baryon interaction from lattice QCD}

\author{\speaker{Takashi Inoue}\\
        Nihon University, College of Bioresource Sciences, 
        Fujisawa 252-0880, Japan\\
        E-mail: \email{inoue.takashi@nihon-u.ac.jp}}

\author{
for HAL QCD Collaboration
}
\author{
\begin{center}
\includegraphics[width=0.33\textwidth]{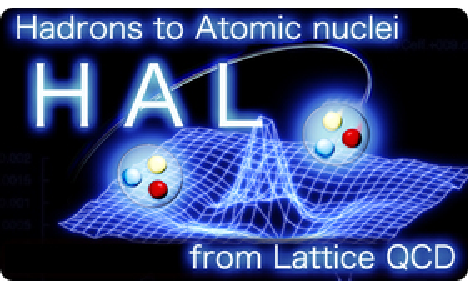}
\end{center}
}

\abstract{
We investigate baryon-baryon ($BB$) interactions 
in the 3-flavor full QCD  simulations with degenerate quark masses for all flavors. 
The $BB$ potentials in the orbital $S$-wave are extracted from the Nambu-Bethe-Salpeter 
wave functions measured on the lattice.
We observe strong flavor-spin dependences of the $BB$ potentials at short distances.
In particular, a strong repulsive core exists in the
flavor-octet and spin-singlet channel
(the ${\bf 8}_s$ representation),
while an attractive core appears in the flavor singlet channel (the  ${\bf 1}$ representation).
We discuss the relation of such flavor-spin dependence with the Pauli exclusion principle 
at the quark level.  The possible existence of an $H$-dibaryon resonance above the $\Lambda\Lambda$ 
threshold is also discussed.
}

\FullConference{The XXVIII International Symposium on Lattice Field Theory, Lattice2010\\
		June 14-19, 2010\\
		Villasimius, Italy}

\begin{document}

\section{Introduction}%
The generalized nuclear force, which includes not only the nucleon-nucleon ($NN$) interaction 
but also hyperon-nucleon ($YN$) and hyperon-hyperon ($YY$) interactions,
has been one of the central topics in hadron and nuclear physics.
Experimental studies on the ordinary and hyper nuclei as well as the 
observational studies of the neutron stars and supernova explosions are
deeply related to the physics of the generalized nuclear force \cite{Hashimoto:2006aw}.

Concerning the $NN$ interaction, significant number of
scattering data have been accumulated and are parametrized
in terms of the phase shifts or of the phenomenological potentials \cite{Machleidt:2001rw}.
On the other hand, the properties of the $YN$ and $YY$ interactions are hardly known even today
due to the lack of high precision $YN$ and $YY$ scattering data.
The approximate flavor $SU(3)$ symmetry does not necessarily create
a bridge between the nuclear force and the hyperon forces because
there are six independent flavor-channels for the scatterings between octet-baryons.
Although there have been previous theoretical attempts to fill the gap on the basis
of the phenomenological quark models \cite{Oka:2000wj} and
of the one-boson-exchange models \cite{Rijken:2010zz},
it is much more desirable to study the generalized nuclear force from the first principle QCD. 

Recently a method to extract the interaction potential from lattice QCD simulations
has been proposed and applied to the $NN$ system \cite{Ishii:2006ec}, 
and $\Xi N$ and $\Lambda N$ systems \cite{Nemura:2010nh} as well.
This paper is our exploratory attempt to unravel the flavor-spin structure of the $BB$ interaction
on the basis of the lattice QCD simulations\cite{Inoue:2010hs}.
To capture the essential structure, we take the finite but degenerated u, d, s-quark masses,
so that the system is in the flavor $SU(3)$ limit.
In the limit a convenient base set exist to look into the structure.
Moreover, $SU(3)$ limit result is easy to compare with effective models
and useful to pin down physical origin of the $BB$ interaction. 

In the next section, we introduce the flavor $SU(3)$ limit and the formalism we employ.
In section 3, we explain a setup of our numerical simulations.
In section 4, we show the resulting $BB$ potentials and discuss their implications. 
The last section is devoted to summary and outlook.

\section{\label{sec:su3}Baryon-baryon potentials in the flavor $SU(3)$ limit}
 
In the flavor $SU(3)$ symmetric limit, 
all baryons and baryonic systems are strictly classified into multiplets
which serve irreducible representation of the flavor $SU(3)$ group.
For example, two-octet-baryon systems are classified as
\begin{equation}
 {\bf 8} \otimes {\bf 8} 
 = \underbrace{{\bf 27} \oplus {\bf 8}_s \oplus {\bf 1}}_{\mbox{symmetric}} ~ 
  \oplus \underbrace{{\bf 10}^* \oplus {\bf 10} \oplus {\bf 8}_a}_{\mbox{antisymmetric}} 
\end{equation}
where ``symmetric" and ``antisymmetric" stand for the symmetry under the
flavor exchange of two baryons.
The Pauli principle between two baryons imposes,
for the system with even (odd) orbital angular momentum,
${\bf 27}$, ${\bf 8}_s$ and ${\bf 1}$ to be spin singlet (triplet),
while ${\bf 10}^*$, ${\bf 10}$ and ${\bf 8}_a$ to be spin triplet (singlet).
These six states provide good basis to describe $BB$ interaction.
In particular, for $S$-wave, all off-diagonal interactions vanish
and six diagonal one remain. The corresponding potentials are
 \begin{eqnarray}
^1S_0 \ &:& \  V^{({\bf 27})}(r), \ V^{({\bf 8}_s)}(r), \ V^{({\bf 1})}(r), 
 \nonumber
\\ 
^3S_1 \ &:& \ V^{({\bf 10}^*)}(r), \ V^{({\bf 10})}(r), \ V^{({\bf 8}_a)}(r) ~.
\label{eqn:sixpot}
\end{eqnarray}
These six potentials contain the essential structure of $S$-wave $BB$ interaction.
In fact, any potential among octet-baryons, including both the diagonal one ($B_1B_2 \rightarrow B_1 B_2)$ and  
the off-diagonal one ($B_1B_2 \rightarrow B_3 B_4$), is obtained by suitable combinations
of these six potentials. 

According to Refs.~\cite{Ishii:2006ec},
non-local but energy-independent potential can be defined and extracted in lattice QCD, 
from the Nambu-Bethe-Salpeter (NBS) wave function $\phi(\vec r\,)$ and the energy $E$, 
through the Schr\"odinger type equation.
As the leading order of the derivative expansion of the non-local potential,
a local potential is obtained by 
\begin{equation}
 V^{(\alpha)}(\vec r\,) = \frac{1}{2\mu}
 \frac{\nabla^2 \phi^{(\alpha)}(\vec r\,)}{\phi^{(\alpha)}(\vec r\,)} + E^{(\alpha)}
 \label{eqn:vr}
\end{equation}
where $\mu$ is the reduced mass of the system.
It is shown that the leading local potential well dominate
the full potential at low energy region \cite{Murano:2010tc}.
We use eq. (\ref{eqn:vr}) and obtain the six potentials in eq. (\ref{eqn:sixpot}).
For $^3S_1$ states, we do not decompose central and tensor potential in this study,
and hence the resulting potential is the one so called ``effective central'' potential.

In the lattice QCD simulations, the NBS wave function $\phi^{(\alpha)}(\vec r\,)$
for the smallest energy is extracted from the four-point function as
\begin{eqnarray}
 G_4^{(\alpha)}(t-t_{0},\vec r\,) 
 =  \mate<0|(BB)^{(\alpha)} (t,\vec r\,)\  \overline{(BB)}^{(\alpha)}(t_0)|0> 
 \propto   \phi^{(\alpha)}(\vec r\,) e^{- W(t-t_0)}
\end{eqnarray}
for $t-t_0 \gg 1$, where $\overline{(BB)}^{(\alpha)}(t_0)$ is
a wall source operator at time $t_0$ to create two-baryon states in $\alpha$-plet,
while $(BB)^{(\alpha)} (t,\vec r\,)$ is the sink operator at time $t$ to annihilate two-baryon states.
These two-baryon operators can be constructed with product of 
the baryon composite field operator and the $SU(3)$ Clebsch-Gordan (CG) coefficients. 

\section{Numerical simulations}

\begin{table}[t]
\caption{\label{tbl:lattice} Summary of lattice parameters and hadron masses.
 For details about the action and the lattice spacing $a$,
 visit the official website of CP-PACS and JLQCD Collaborations\cite{CPPACS-JLQCD}.}
 \begin{tabular}{c|c|c|c|c|c|c|c}
   \hline \hline
    lattice  & $\beta$ & ~$a$ [fm] ~& ~L [fm] ~  
             & $\kappa_{uds}$  & ~$m_{\rm ps}$ [MeV]~ & ~ $m_{B}$ [MeV]~ & ~$N_{\mbox{cfg}}$~ \\
   \hline 
   \multirow{2}{*}{$16^3 \times 32$} & ~\multirow{2}{*}{1.83}~
                                     & \multirow{2}{*}{0.121(2)} & \multirow{2}{*}{1.93(3)}
          & ~0.13710~ & 1014(1) & 2026(3) & 700 \\
    & & & & ~0.13760~ & ~835(1) & 1752(3) & 800 \\
   \hline \hline
 \end{tabular}
\end{table}

We utilize two gauge configuration sets 
provided by the Japan Lattice Data Grid (JLDG) and International Lattice Data Grid (ILDG) \cite{JLDG/ILDG},
which are generated by CP-PACS and JLQCD Collaborations \cite{CPPACS-JLQCD},
with the renormalization-group-improved Iwasaki gauge action
and the non-perturbatively $O(a)$-improved Wilson quark action. 
The lattice parameters and hadron masses are summarized in Table \ref{tbl:lattice}.
Those hopping parameters $\kappa_{uds}$ reside around the physical strange quark mass
region in 2+1 flavor QCD at the same $\beta$\cite{CPPACS-JLQCD}.

We calculate quark propagators for the spatial wall source at $t_0$ 
with the Dirichlet boundary condition in the temporal direction at $t=t_0+16$.
In order to enhance the signal, we use all 32 time slices on each configuration as $t_0$,
and take the average over forward and backward propagations in time. 
We estimate the statistical errors by the jackknife method.
All these numerical computations have been done
at the KEK supercomputer system, Blue Gene/L and SR11000.

\section{Results and their implications}

\subsection{$BB$ potentials in flavor basis}


\begin{figure}[tp]
 \includegraphics[width=0.49\textwidth]{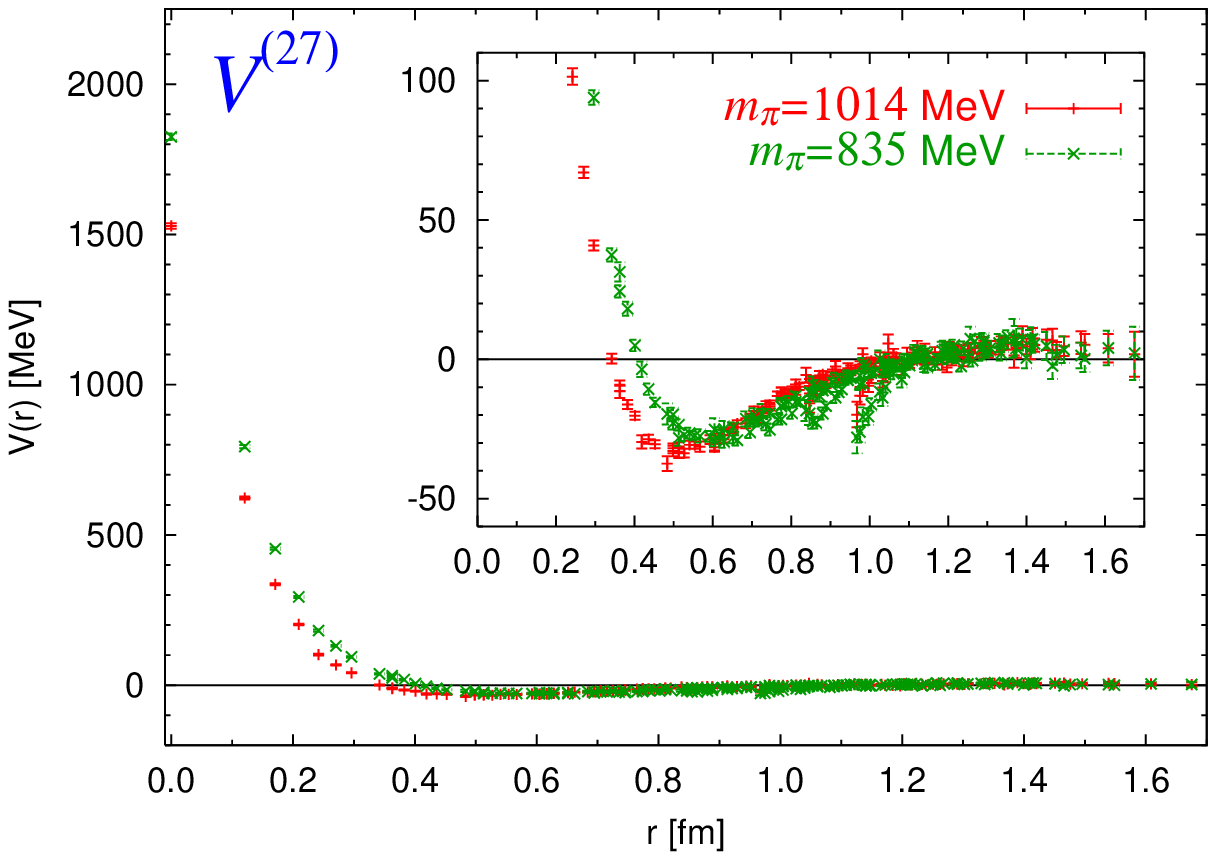}\hfill
 \includegraphics[width=0.49\textwidth]{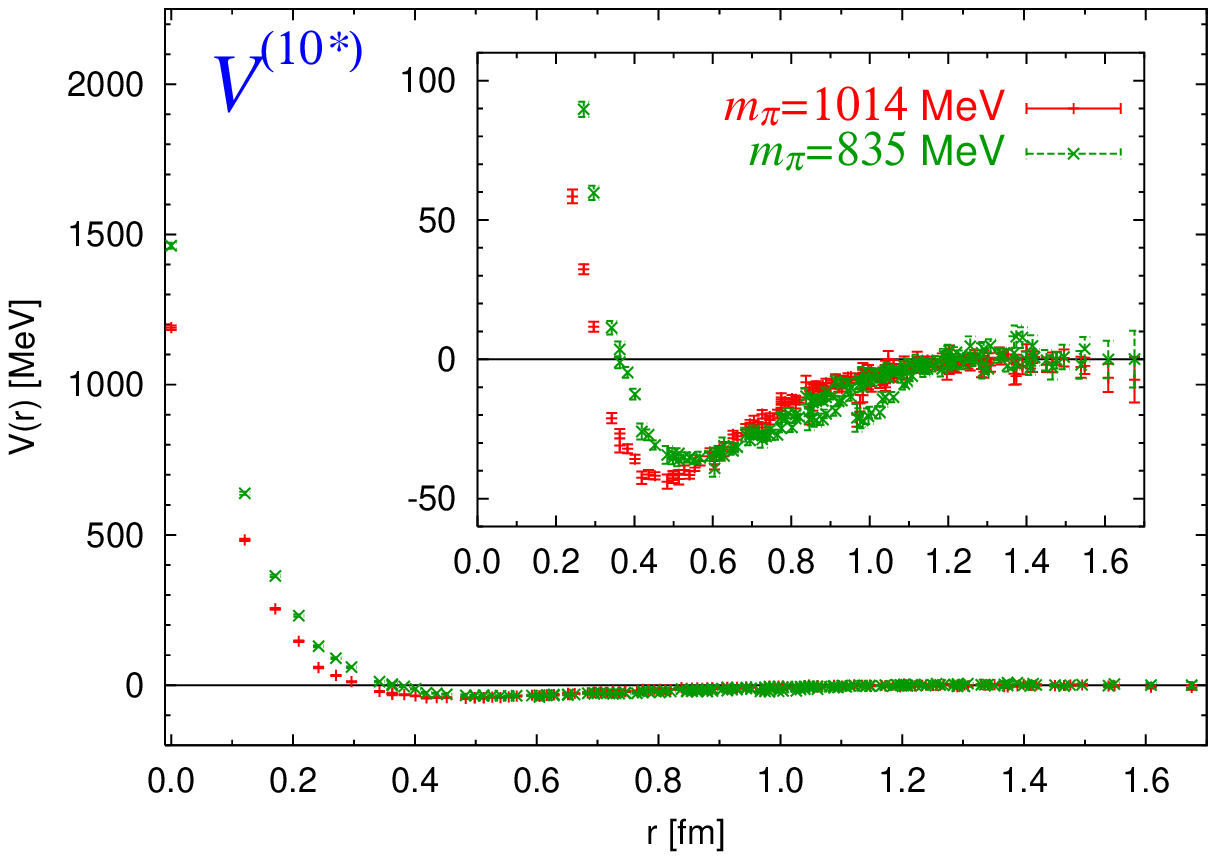}
\caption{\label{fig:V27andV10s}
  Lattice QCD extracted potentials $V^{({\bf 27})}$ in the left and $V^{({\bf 10}^{*})}$ in the right.
 } 
\end{figure}

\begin{figure}[tp]
 \includegraphics[width=0.49\textwidth]{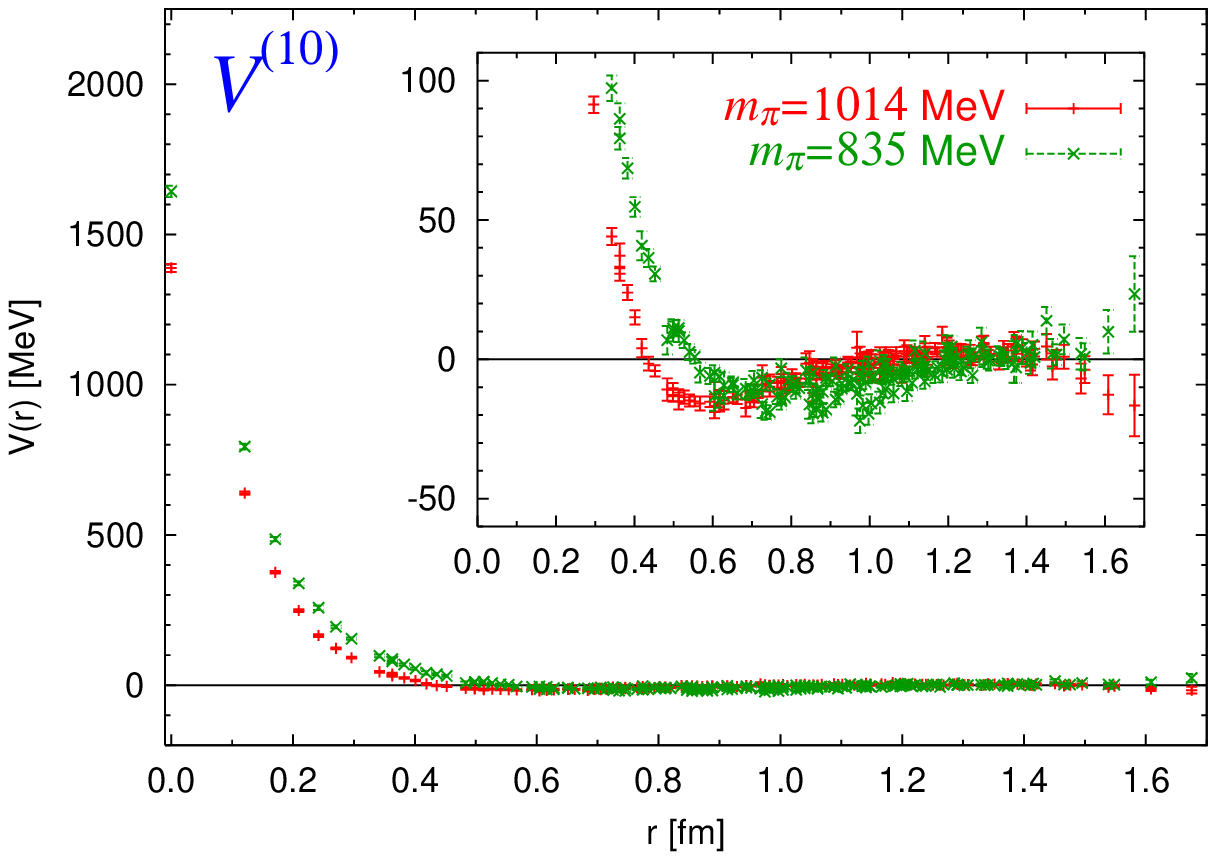}\hfill
 \includegraphics[width=0.49\textwidth]{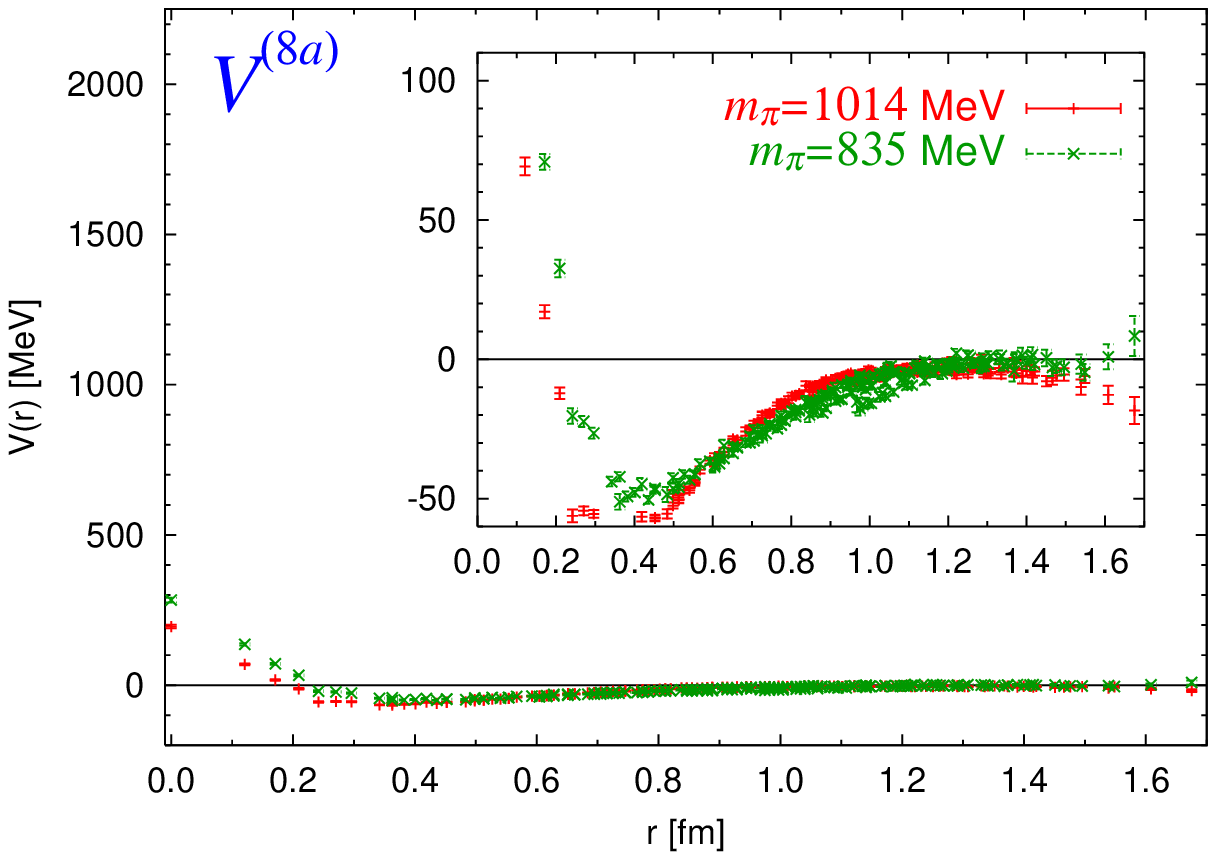}
\caption{\label{fig:V10andV8a}
  Lattice QCD extracted potentials $V^{({\bf 10})}$ in the left and $V^{({\bf 8}_a)}$ in the right.
 } 
\end{figure}

\begin{figure}[tp]
 \includegraphics[width=0.49\textwidth]{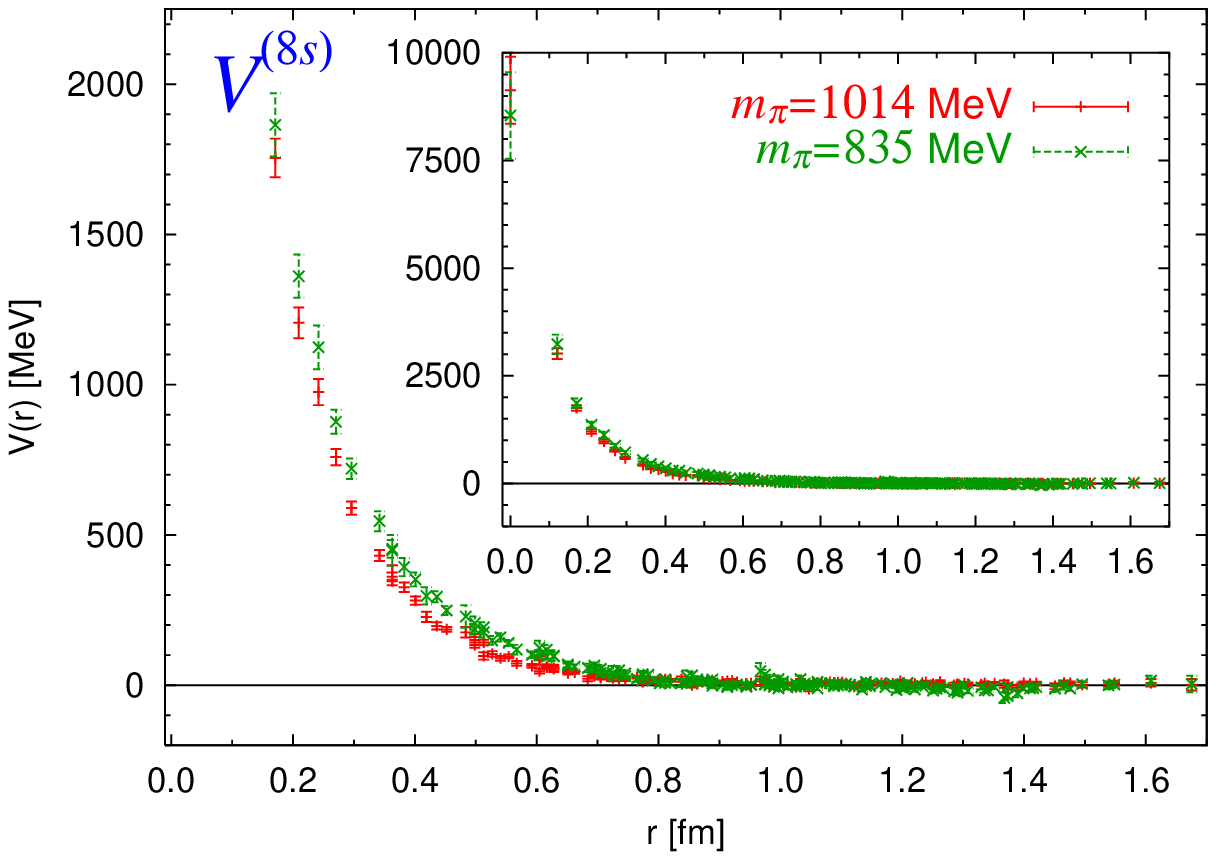}\hfill
 \includegraphics[width=0.49\textwidth]{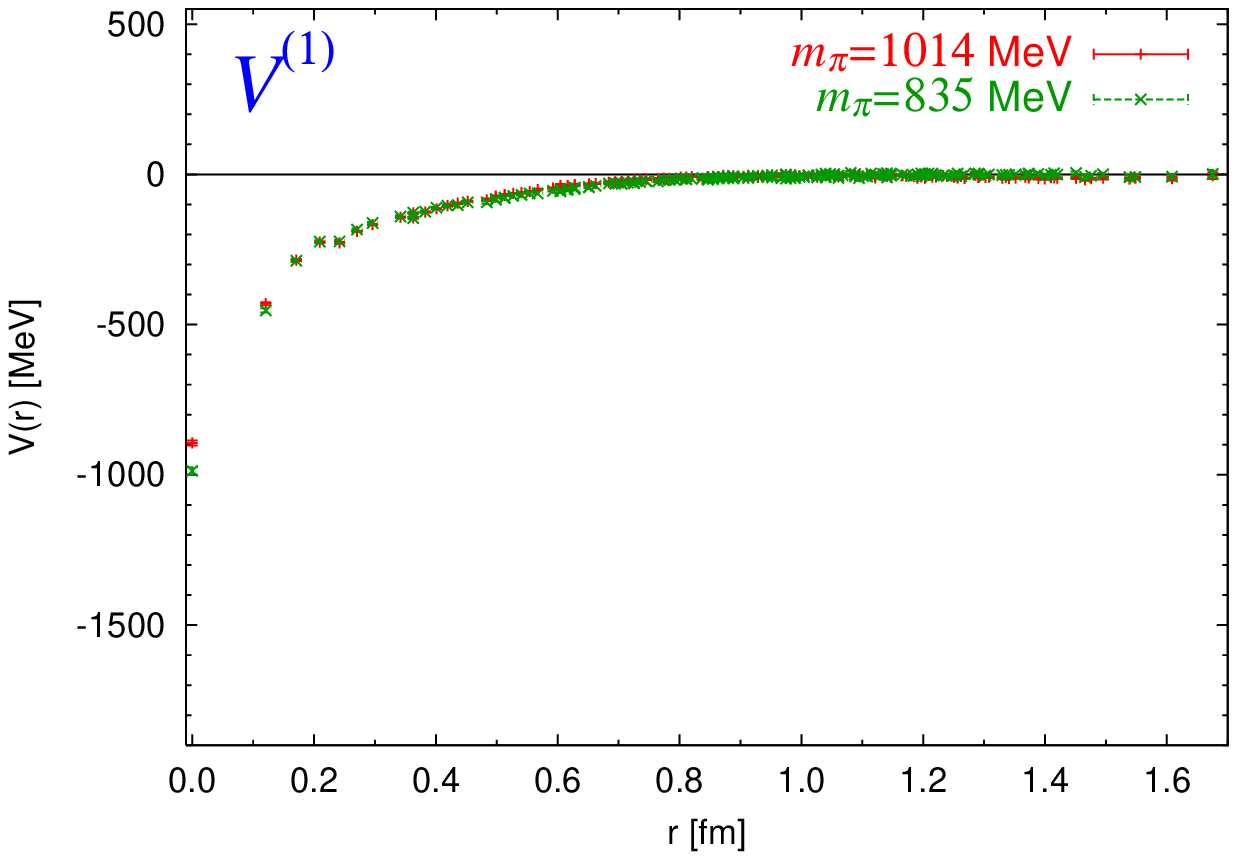}
\caption{\label{fig:V8sandV1}
  Lattice QCD extracted potentials $V^{({\bf 8}_s)}$ in the left and $V^{({\bf 1})}$ in the right.
 } 
\end{figure}

Figure~\ref{fig:V27andV10s} to \ref{fig:V8sandV1} show the obtained six $BB$ potentials.
Red bars (green crosses) correspond to the pion mass of 1014 MeV (835 MeV).
Since the effective mass of the four-point function in each channel shows 
a plateau at $t-t_0 \ge 10$, we use data at $t-t_0=10$ exclusively
to extract both NBS wave functions and $BB$ potentials throughout this paper. 
We have estimated the energies contribution
(the second term on the right-hand side of Eq.~(\ref{eqn:vr})) 
under the condition that Eq.~(\ref{eqn:vr}) approaches zero at large $r$ 
(the method called ``from $V$" in Ref.~\cite{Ishii:2006ec}).  
We find $E^{({\bf 27})}  \simeq -5$ MeV,\,  $E^{({\bf 8}_s)} \simeq 25$ MeV,\,
$E^{({\bf 1})}  \simeq -30$ MeV,\, $E^{({\bf 10}^*)} \simeq -10$ MeV,\,
$E^{({\bf 10})} \simeq  0$ MeV and $E^{({\bf 8}_a)} \simeq -15$ MeV.
Since the energy contribution is small compared to the Laplacian term
(the first term on the right-hand side of Eq.~(\ref{eqn:vr})) at a short distance,
we have not attempted to extract a precise value of  $E^{(\alpha)}$ in this study.

Figure~\ref{fig:V27andV10s} shows potentials $V^{({\bf 27})}$ and  $V^{({\bf 10}^*)}$.
Because a symmetric $NN$ belongs to ${\bf 27}$ and anti-symmetric $NN$ belongs to ${\bf 10^*}$
in the $SU(3)$ limit, $V^{({\bf 27})}$ and $V^{({\bf 10}^*)}$ can be regarded
as the $SU(3)$ limit of the nuclear force potential in $^1S_0$ and $^3S_1$, respectively.
Both potential have a repulsive core at a short distance and an attractive pocket at around 0.6 fm.
These qualitative features agree with our previous results found in the $NN$ system
in quenched approximation with lighter quark mass\cite{Ishii:2006ec}.
This gives good consistency check.
Furthermore, these feature qualitatively agree with well known property of 
the phenomenological $NN$ potentials constructed form data.
This shows efficiency of the present method.

Figure~\ref{fig:V10andV8a} shows potentials $V^{({\bf 10})}$ and $V^{({\bf 8}_a)}$.
Since anti-symmetric $N \Sigma$($I$=3/2) belongs to ${\bf 10}$ in the $SU(3)$ limit,  
one can regard $V^{({\bf 10})}$ as the $SU(3)$ limit of $^3S_1$ $N \Sigma$($I$=3/2) potential, for example.
Similarly, $V^{({\bf 8}_a)}$ is the $SU(3)$ limit of $^3S_1$ $N \Xi$($I$=0) potential.
As you see, $V^{({\bf 10})}$ has a stronger repulsive core and a much shallower attractive pocket
than  $V^{({\bf 27},{\bf 10}^*)}$. 
On the other hand, $V^{({\bf 8}_a)}$ has a very weak repulsive core and a deeper attractive pocket
compared to $V^{({\bf 27},{\bf 10}^*)}$. 
This is exactly a part of the essential structure of $BB$ interaction we looked for.
It is interesting to point out that the above features were foreseen in a simple quark model, 
where effect of quark Pauli blocking and one-gluon-exchange interaction
are evaluated systematically using the $SU(6)$ quark wave function of baryon\cite{Oka:2000wj}.
We come back this point later.

Figure~\ref{fig:V8sandV1} shows potentials $V^{({\bf 8s})}$ and $V^{({\bf 1)}}$.
These potentials cannot be interpreted as an $SU(3)$ limit of some potential between two baryons,
but are always superpositions of them.
First, we see that $V^{({\bf 8}_s)}$ has a very strong repulsive core.
It is strongest among all the channels.
On the other hand, $V^{({\bf 1})}$ shows no repulsion at a short distance but attraction instead.
It is in contrast to other cases.
Now, the essential flavor-spin structure of the S-wave $BB$ interaction is revealed completely.

As the case of $V^{({\bf 10})}$ and $V^{({\bf 8}_a)}$,
the above features of $V^{({\bf 8}_s)}$ and $V^{({\bf 1})}$ were
also foreseen in simple quark models\cite{Oka:2000wj,Jaffe:1976yi}.
In particular, $V^{({\bf 8}_s)}$ in quark models becomes strongly repulsive at a short distance
since the six quarks cannot occupy the same orbital state due to quark Pauli blocking. 
On the other hand, in the ${\bf 1}$ channel six quarks do not suffer from the quark Pauli blocking at all,
and the potential can become attractive due to short-range gluon exchange. 
In total, simple quark models predicted the essential feature of $BB$ interaction correctly.
This fact suggests that the quark Pauli blocking plays an essential role
in the repulsion in $BB$ system.  

\subsection{$BB$ potentials in baryon basis}

With explicit flavor $SU(3)$ breaking,
the baryon basis such as $NN$ and $\Lambda N$ are more suitable to describe $BB$ interactions
than the flavor basis.
Generally, interaction potential takes matrix form in coupled channels in the baryon basis.
In the $SU(3)$ limit, the baryon-base potentials 
can be obtained by a unitary rotation of flavor-base potentials as
$V_{ij}(r) = \sum_{\alpha} U_{i\alpha} \,  V^{(\alpha)}(r) \, U^{\dagger}_{\alpha j}$
where $U$ is a unitary matrix made of CG coefficients, 
which rotates the flavor basis to the baryon basis.

\begin{figure}[t]
 \includegraphics[width=0.49\textwidth]{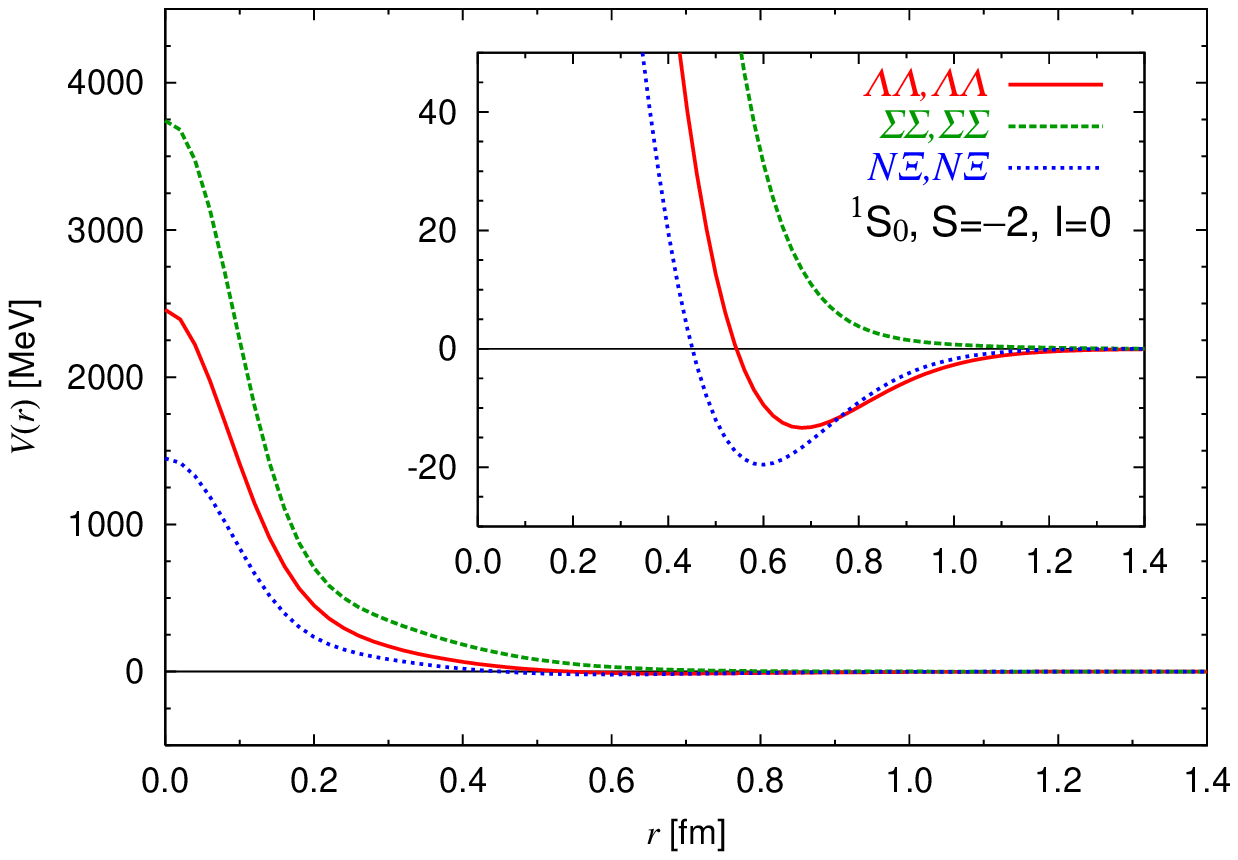}\hfill
 \includegraphics[width=0.49\textwidth]{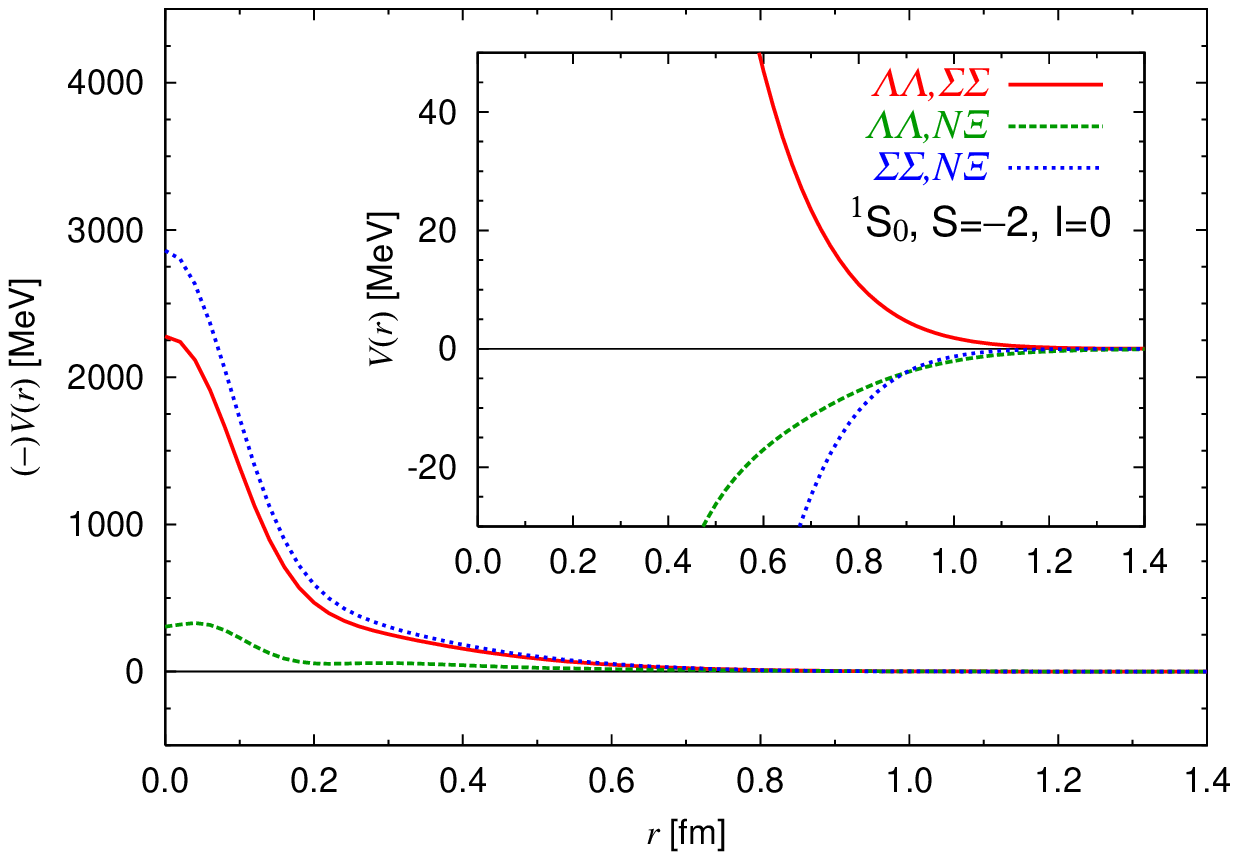}
 \caption{\label{fig:pot_lamlam}
  $BB$ potentials in baryon basis for $S$=$-$2, $I$=0, $^1S_0$ sector. 
  Three diagonal (off-diagonal) potentials are shown in the left (right) panel.
  Phases of off-diagonal ones in the right panel are arranged in a zoom-out plot. 
  Their true signs are shown in the insert.
 }
\end{figure}

In Figure~\ref{fig:pot_lamlam}, we show the potentials for $^1S_0$, $S$=$-$2, $I$=0 sector as characteristic examples.
To obtain $V_{ij}(r)$, we have used a simple analytic function with five parameters
fitted to data of $V^{(\alpha)}$.
The left panel of  Fig.~\ref{fig:pot_lamlam} shows the diagonal part of the potentials.
We see that $\Lambda\Lambda$ and $N\Xi$($I$=0) channels are attractive
while $\Sigma\Sigma$($I$=0) channel is repulsive.
The attraction in the ${\bf 1}$ channel is most reflected in the $N\Xi$($I$=0) channel, 
while the strong repulsion in the ${\bf 8}_s$ channel is most reflected in the $\Sigma\Sigma$($I$=0),
owing to its largest CG coefficient among the three channels.
Nevertheless, all three diagonal potentials have a repulsive core originating from the ${\bf 8}_s$ component.
The right panel of Fig.~\ref{fig:pot_lamlam} shows the off-diagonal parts of the potential matrix,
which are comparable in magnitude to the diagonal ones.
Since the off-diagonal parts are not negligible in this sector,
full coupled channel analysis is necessary to study the observables.

Similarly, we have obtained potentials for all other sectors with different
strangeness, iso-spin and spin, and find their features\cite{Inoue:2010hs}.
Since those features may remain in the flavor-$SU(3)$-broken physical world essentially,
we expect our result provide useful hints on the behavior of hyperons ($\Lambda$, $\Sigma$ and $\Xi$)
in hypernuclei and in neutron stars \cite{Hashimoto:2006aw}.

\section{Summary and outlook}

We have performed 3-flavor full QCD simulations
to study the general features of the $BB$ interaction in the flavor $SU(3)$ limit.
From the NBS wave function measured on the lattice, 
we have extracted all six independent potentials in the $S$-wave 
at the leading order of the derivative expansion.
We have found strong flavor-spin dependence of the $BB$ interactions. 
It turns out there exist variety of $BB$ interaction in u, d, s-quark three-flavor world,
in contrast to u, d-quark two-flavor world.
In particular, $V^{({\bf 8}_s)}$ has a very strong repulsive core at a short distance, 
while $V^{({\bf 1})}$ is attractive at all distances. 
These features are consistent with the Pauli blocking effect among quarks
previously studied in phenomenological quark models. 
This indicate that the quark Pauli blocking is essential for the repulsion in $BB$ interactions
at a short distance.

In the flavor singlet $BB$ channel, a bound state so called $H$-dibaryon
is expected in phenomenological quark models\cite{Jaffe:1976yi}.
The present lattice QCD data are insufficient to derive a definite conclusion on the $H$-dibaryon,
because of a single and small lattice volume.
Now we are performing lattice QCD simulations at flavor-$SU(3)$-point in larger volume. 
Nevertheless, using the potential $V^{({\bf 1})}(r)$ and solving the Schr\"odinger equation,
a shallow bound state is found, suggesting a possibility of a bound $H$-dibaryon
with baryon components $\Lambda\Lambda:\Sigma\Sigma:\Xi N =-\sqrt{1}:\sqrt{3}:\sqrt{4}$,
in the $SU(3)$ limit world with heavy quark.
In the real world, flavor $SU(3)$ breaking is considerable and 
the thresholds of the $S$=$-$2, $I$=0 sector are located in order $\Lambda\Lambda < \Xi N < \Sigma\Sigma$.
In our test study of solving the Schr\"odinger equation
with small baryon mass differences from a 2+1 flavor lattice QCD simulation,
a resonance state is found at an energy between $\Lambda\Lambda$ and $\Xi N$ threshold.
If such a resonance exists in nature, it may explain the enhancement just above the $\Lambda \Lambda$ threshold
recently reported in the KEK experiment \cite{Yoon:2007aq}.
However, further investigations in both theory and experiment
are necessary to derive a definite conclusion.


\section*{Acknowledgements}
The author thank the CPS for their lattice QCD simulation code \cite{CPS}, 
the CP-PACS and JLQCD Collaborations for their gauge configurations \cite{CPPACS-JLQCD}, 
which are provided through the JLDG / ILDG \cite{JLDG/ILDG}. 
This research is supported in part by 
Grants-in-Aid for Scientific Research on Innovative Areas (No.~2004:20105001, 20105003)
and the Large Scale Simulation Program No.~09-23 (FY2009) of 
the High Energy Accelerator Research Organization (KEK).

\end{document}